\documentclass[aps,prl,groupedaddress,amsmath,amssymb,twocolumn]{revtex4}

\usepackage{graphicx}

\begin{document}
\title{Slow-light effect in dual-periodic photonic lattice}
\author{Alexey G. Yamilov\footnote{Electronic address: yamilov@umr.edu}, Mark R. Herrera}
\affiliation{Department of Physics, University of Missouri--Rolla, Rolla, MO 65409\footnote{Effective Jan. 1, 2008, UMR will become Missouri University of Science and Technology (Missouri S\&T)}}
\author{Massimo F. Bertino}
\affiliation{Department of Physics, Virginia Commonwealth University, Richmond, VA 23824}
\date{\today}

\begin{abstract}
We present analytical and numerical study of photonic lattice with short- and long-range harmonic modulations of the refractive index. Such structures can be prepared experimentally with holographic photolithography. In the spectral region of photonic bandgap of the underlying single-periodic crystal, we observe a series of bands with anomalously small dispersion. The related slow-light effect is attributed to the long-range modulation in the photonic lattice, that leads to formation of an array of evanescently-coupled high-$Q$ cavities. The band structure of the lattice is studied with several techniques: (i) transfer matrix approach; (ii) analysis of resonant coupling in process of band folding; (iii) effective medium approach based on coupled-mode theory; and (iv) Bogolyubov-Mitropolsky approach. The latter method, commonly used in the studies of nonlinear oscillators, was employed to investigate the behavior of the eigen-function envelopes and the  band structure of our dual-periodic photonic lattice. We show that reliable results can be obtained even in the case of large refractive index modulation. \\
\end{abstract}
\maketitle


\section {Introduction}

One of the attractive features of photonic crystals (PhCs) is the possibility to alter the dispersion of the electromagnetic waves\cite{phc}. ``Slow light''\cite{milonni} is the property that attracted a great deal of practical interest as it leads to low-threshold lasing\cite{nojima,sakoda,susa}, pulse delay\cite{yariv_delay_line,vlasov_delay_line}, optical memories \cite{yariv_opn}, and to enhanced nonlinear interactions \cite{joannopoulos_nonlinearity_in_phc,yariv_shg}. There are several approaches to slowing down the light in PhC structures:

(i) At the edge of a bandgap, the dispersion (group velocity) $v_g=d\omega/dK$ becomes zero. This property has been extensively studied and used in practice for control of spontaneous emission \cite{yablonovitch_spontaneous_emission} and gain enhancement in lasers \cite{nojima,sakoda,susa}. In the vicinity of the band-edge the group velocity strongly depends on frequency leading to strong distortion of a pulsed signal, rendering it impractical for {\it e.g.} information processing applications.

(ii) Unlike one-dimensional PhC, in two- and three-dimensional structures the dispersion of the high-frequency bands becomes smaller over the finite region of frequencies. This is due to the increased density of states which leads to band-interaction and flattening of the photonic bands \cite{lopez,mike_high_order}. These high-frequency features, however, are of limited use in practice because they allow little control, are not spectrally isolated, and are usually susceptible to disorder \cite{dorado}. 

(iii) Based on the Coupled Resonator Optical Waveguide (CROW)\cite{stefanou,yariv_crow_proposal1,yariv_crow_proposal2,yariv_opn} idea, the periodic array of weakly coupled optical resonators made in a photonic crystal\cite{1d_crow1,1d_crow2,vuckovic_low_threshold_phc_laser} mitigates the limitations in approaches (i-ii). Hybridization of high-Q resonances due to periodically positioned structural defects can form a defect-band in the spectral region of the photonic bandgap:
\begin{equation}
\omega (q)=\Omega [1+\kappa \cos (qa)]
\label{dispersion}
\end{equation}
As a result, the group velocity is reduced over a substantial range of frequencies, whereas the band characteristics can be controlled via the parameters of the structural defects.

A periodic arrangement of structural defects in the photonic crystal creates a {\it photonic super-crystal} (PhSC) with short-range quasi-periodicity on the scale of the lattice constant and long-range periodicity on the defect separation scale \cite{1d_dual_theory,1d_dual_experiment,1d_dual_butterfly,1d_dual_chinese_group,1d_transfer_matrix1,1d_transfer_matrix2,bragg_grating}. We recently proposed PhSC with dual-{\it harmonic} modulation of refractive index\cite{our_ol_crow} that can be fabricated\cite{bertino_apl2004} via a single-step interference photolithography technique. Because all resonators are produced at once, our design minimizes fabrication error margin and ensures the large-scale periodicity which is essential for hybridization of the resonances of individual cavities in an experiment. 

The purpose of this work is to theoretically investigate the optical properties of one-dimensional PhSC with a combination of analytical and numerical techniques. We consider systems where the dielectric function is given by
\begin{equation}
\varepsilon(x)=\varepsilon_{0}+\frac{\Delta\varepsilon /2}{1+\gamma}\left[1+\gamma\cos\left( 2\pi x/L\right) \right] \left[ 1+\cos\left( 2\pi x/a \right) \right].
\label{epsilon}
\end{equation}
Here $\varepsilon_0$ is the background dielectric constant. The amplitude of the short-range (on scale $a$) modulation gradually changes from $\Delta\varepsilon\times (1-\gamma)/(1+\gamma)$ to $\Delta\varepsilon$. $L$ sets the scale  of the long-range modulation and we will concentrate on $L/a\gg1$ limit. An interesting situation may arise when $L/a$ is non-integer. The spectrum of such structures exhibits the characteristic ``butterfly'' structure\cite{1d_dual_butterfly}, similar to the spectrum of an electron in a solid with an applied magnetic field\cite{bloch_Hfield_butterfly}. We, however, will limit our consideration to commensurate lattices where $L/a=N$ is a large integer. 

The paper is organized as follows. In Section 2 we discuss the possibility of reproducing the proposed structures in holographic photolithography experiments. In Section 3 we investigate the optical spectra of finite and the photonic band structure of the infinite PhSC with transfer matrix technique. In Section 4 we study photonic band folding for moderate values of $N$ and discuss the onset of the formation of photonic bands with small dispersion. In Section 5 we apply perturbative, ``effective medium'', approximation to our dual periodic crystals with large $N$. Using separation of scales $L\gg a$ technique from nonlinear oscillator theory, in Section 6 we derive a set of first order differential equations that describe our problem on the language of eigen-function envelope. We conclude with the summary Section 7.

\section {An experimental realization of PhSC}

In our previous work \cite{our_ol_crow} we proposed to employ holographic photolithography to fabricate a modulated photonic crystal, PhSC. The experimental quantum dot UV photolithography (QDPL) technique -- a bottom-up technique that allows to synthesize quantum dots and other nanoparticles in selected regions of porous matrices \cite{bertino_apl2004} -- is suitable for this purpose. The change of the dielectric function is related to the local density of quantum dots and it is proportional to the local intensity of the electromagnetic wave used when exposing the sample. We studied a setup where four S-polarized laser beams are defined by  
\begin{equation}
\left[\begin{array}{c}
{\bf q}_{L1},E_{L1}\\
{\bf q}_{L2},E_{L2}\\
{\bf q}_{R1},E_{R1}\\
{\bf q}_{R2},E_{R2}\\
\end{array}\right]=
\left[\begin{array}{c}
k_0\{-\sin (\theta_1),0,\cos(\theta_1)\},E_1\\
k_0\{-\sin (\theta_2),0,\cos(\theta_2)\},E_2\\
k_0\{ \sin (\theta_1),0,\cos(\theta_1)\},E_1\\
k_0\{ \sin (\theta_2),0,\cos(\theta_2)\},E_2\\
\end{array}\right].
\label{holography}
\end{equation}

Here ${\bf q}$ and $E$ are  the k-vector and amplitude of the beams respectively. Their interference $E_{tot}(x)\propto \alpha \cos (k_1x)+\beta \cos (k_2x)$ leads to  the following expression for dielectric constant
\begin{equation}
\varepsilon(x)=\varepsilon_0+\Delta\varepsilon\left[ \alpha \cos (k_1x)+\beta \cos (k_2x) \right]^2,
\label{modulation}
\end{equation}
where $\alpha=E_{1}/(E_{1}+E_{2})$, $\beta=E_{2}/(E_{1}+E_{2})$, $k_1=k_0\sin (\theta_1)$ and $k_2=k_0\sin (\theta_2)$. Choosing $k_{1,2}=\pi/a\pm\pi/L$ and $\alpha=\gamma /\left[1+\gamma\right]$ creates PhSC with dual-harmonic modulations, short- and long-range, similar to that in Eq. \ref{epsilon}. Manipulation of the lithographic beams allows for easy control over the structural properties of the resultant PhC: (i) fundamental periodicity $a$ via $k_0$ and $\theta_{1,2}$; (ii) long-range modulation $L$ via $\theta_1-\theta_2$; (iii) depth of the long-range modulation via relative intensity of the beams $E_1/E_2$. Although the exact $\varepsilon(x)$ dependence in Eq. \ref{modulation} differs from that in Eq. \ref{epsilon}, it shows the same spectral composition and modulation property. If the above structure is realized in experiment, the discrepancy is expected to cause only small deviation from the analytical results obtained further in this work, furthermore, it also becomes insignificant in the limit $L\gg a$. From now on, for the reasons of computational convenience we will only consider dielectric function in form of Eq. \ref{epsilon}.

Dual-periodic harmonic modulation of refractive index can be also experimentally realized in optically-induced photorefractive crystals\cite{photorefracive_crystals}. Although, unlike QDPL\cite{bertino_apl2004}, the obtained index contrast is several orders of magnitude less, the superlattices created in the photorefractive materials offer the possibility of dynamical control. Study of dynamical and nonlinear phenomena in dual-periodic lattices presents significant interest. However, it goes beyond the scope of present publication and will not be considered in this work.

\section {Transfer Matrix Analysis}

\begin{figure}
\vskip -0.5cm
\centerline{\rotatebox{0}{\scalebox{0.4}{\includegraphics{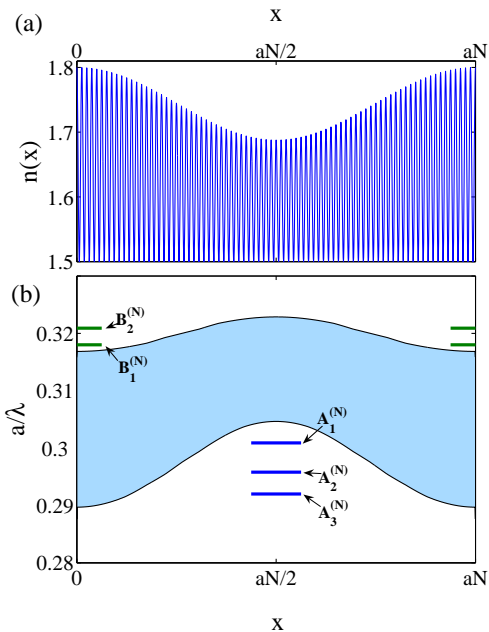}}}}
\vskip -0.5cm
\caption{\label{nofx} (a) Dependence of the index of refraction in dual-periodic photonic crystal as defined by Eq. (2). We used $\varepsilon_0=2.25$, $\Delta\varepsilon=1$, $N=80$ and the modulation parameter $\gamma$ is equal to $0.25$. (b) Local (position-dependent) photonic bandgap diagram for $n(x)$ in (a). $A^{(N)}_i$ and $B^{(N)}_i$ mark the frequencies of the foremost photonic bands on long- and short-wavelength sides of PBG of the corresponding single-periodic crystal.}
\end{figure}

Transmission/reflection spectra of PhSC of finite length as well as the band structure of its infinite counterpart can be obtained numerically via transfer matrix approach. The propagation of the field through a infinitesimal segment of length $dx$ is described by \cite{yeh}
\begin{equation}
\widehat{M}(x,x+dx)= 
\left[ 
\begin{array}{cc}
     \cos(kn(x)dx) & n^{-1}(x)\sin(kn(x)dx)  \\ 
-n(x)\sin(kn(x)dx) &          \cos(kn(x)dx) 
\end{array} 
\right] 
\label{matrix}
\end{equation}
where we assumed that refractive index $n(x)$ does not change appreciably over that distance. The matrix $\widehat{M}(x,x+dx)$ relates the electric field and its spacial derivative $\{E,1/k\ dE/dx\}$ at $x+dx$ and $x$. Total transfer matrix of the finite system is given by the product of the individual matrices 
\begin{equation}
\widehat{M}_{tot}=\displaystyle \prod _{x=0}^{L}\widehat{M}(x,x+dx)
\label{total matrix}
\end{equation}
Since in our case refractive index $n(x)=\varepsilon^{1/2}(x)$, Fig. \ref{nofx}(a), is not a piece-wise constant (as in Ref.\cite{1d_transfer_matrix1,1d_transfer_matrix2}) but a continuous function of coordinate, one has to resort to numerical simulations. In what follows we apply either scattering or periodic boundary conditions to obtain transmission coefficient and Bloch number $q(\omega)$ respectively.

\begin{figure}
\centerline{{\scalebox{0.4}{\includegraphics{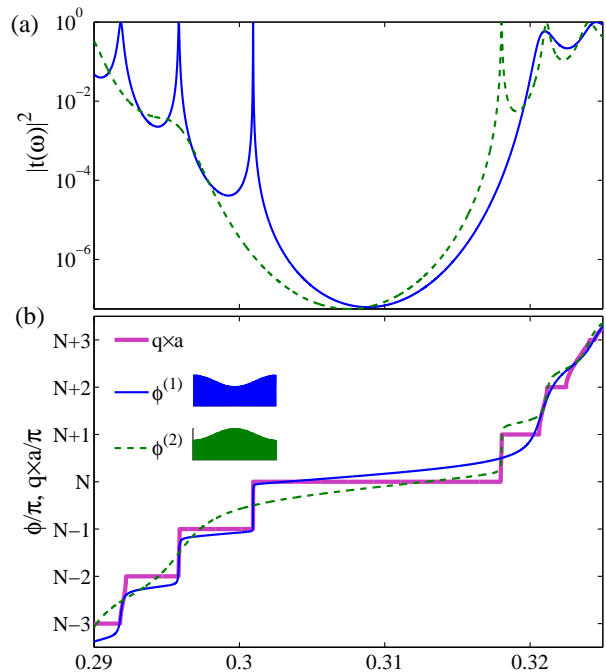}}}}
\vskip -1cm
\caption{\label{transmission} (a) Transmission coefficient through a finite segment of length $L$ (a period) of the periodic super-structure defined in Fig. \ref{nofx}. Solid and dashed lines corresponds to $0<x<Na$ and $-Na/2<x<Na/2$ segments (shown in the inset of panel (b)) respectively. (b) Solid and dashed thin lines plot the corresponding phase of $t(\omega)$. Bold line depicts the Bloch number $q(\omega)\times a$ of the infinite crystal computed using Eq. \ref{q_from_t}.}
\end{figure}

Fig. \ref{transmission}(a) plots transmission coefficient through one period of a dual-periodic system shown in Fig. \ref{nofx}. A series of progressively sharper resonances occur on the lower or upper edge of the spectral gap of underlying single-periodic structure. Whether the peaks occur at lower or upper band edge depends on the particular definition of the unit cell as shown in the inset of Fig. \ref{transmission}(a). One can obtain an insight into this effect by examining the modulation of the spectral position of the ``local'' photonic bandgap with coordinate as shown in Fig. \ref{nofx}(b). This analysis is meaningful on $a\ll \Delta x\ll L\equiv Na$ scale. This condition can be satisfied in our case of slow modulation -- large $N$. At the frequencies such as $A^{(N)}_i$ in Fig. \ref{nofx}(b) the wave propagation is allowed in the vicinity of $x=aN\times (1/2+m)$, whereas $x=aN\times m$ are locally forbidden, $m$ being an integer. When considering $0<x<Na$ segment of the lattice, resonant tunneling via electromagnetic states of the cavity at the geometrical center, $A^{(N)}_i$, leads to the low-frequency peaks in the transmission coefficient -- solid line in Fig. \ref{transmission}(a).On the other hand, transmission through $-Na/2<x<Na/2$ segment exhibits a series of sharp resonances which correspond to tunneling via $B^{(N)}_i$ cavity states in the high-frequency region.

\begin{figure}
\centerline{{\scalebox{0.4}{\includegraphics{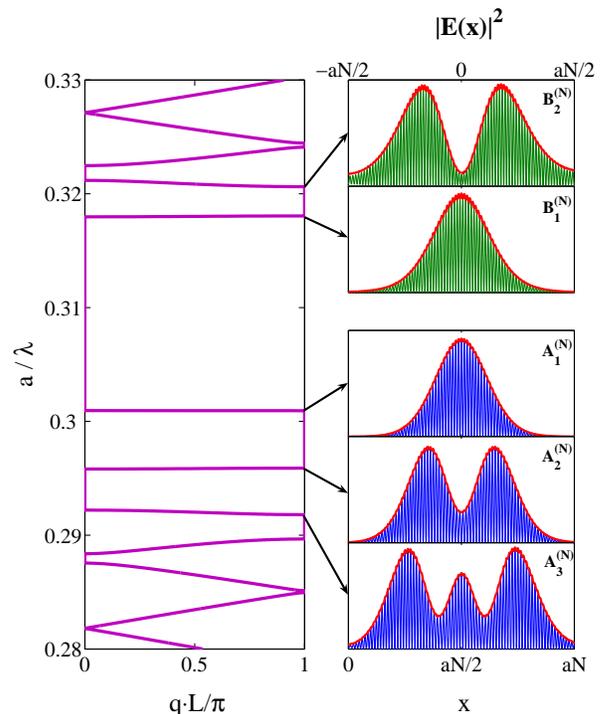}}}}
\vskip -1cm
\caption{\label{intensity} The left panel shows dispersion of PhSC $\omega(q)$ reduced to the first Brillouin zone. The eigen-modes which correspond to the series of flat bands in the vicinity of the parent-bandgap of the single periodic crystal are depicted on the right. Calculations were performed for the structure described in Fig. \ref{nofx}.}
\end{figure}

The transmission coefficient through a finite segment of length $L$ (equal to one period) can be related to the band structure of the corresponding periodic lattice\cite{1d_transfer_matrix2} as
\begin{equation}
\cos(q(\omega)L)=Re[\frac{1}{t(\omega)}]\equiv\frac{1}{|t(\omega)|}\cos(\phi(\omega)),
\label{q_from_t}
\end{equation}  
where we introduced the phase of the transmission coefficient through $t=|t|\exp[i\phi]$. Fig. \ref{intensity} shows that hybridization of the cavity resonances considered above leads to formation of the flat photonic bands. Their low dispersion, small group velocity, may be exploited\cite{our_ol_crow} for practical applications. 

In the vicinity of an isolated transmission resonance, $t(\omega)$ is given as the Lorentzian
\begin{equation}
t(\omega)=\frac{(-1)^{N} (\Gamma / 2)}{i (\Gamma /2)-(\omega-\omega_{0})}
\label{lorentzian}
\end{equation}
where $\Gamma$ is the full width at half maximum (FWHM) of the resonance and $\omega_{0}$ is the resonant frequency. Substitution of Eq. \ref{lorentzian} into Eq. \ref{q_from_t} gives the flat band described by
\begin{equation}
\omega(q)=\omega_{0}[1 \pm \kappa \cos(qa_{L})]
\label{crow_band}
\end{equation}
where $\kappa=\displaystyle\frac{\Gamma}{2 \omega_{0}}=\displaystyle\frac{1}{Q} \ll 1$, and $Q$ is the cavity $Q$-factor. Thus, the decrease of group velocity in the PhSC is directly related to the increase of confinement and decrease of the coupling between the neighboring cavities. In our PhSC both these factors are described by the same parameter -- the cavity $Q$-factor. In a single-periodic PhC of finite length, the $Q$-factor of a band-edge mode depends on the system size. Comparing Figs. \ref{nofx}(b),\ref{intensity} one can see that $A^{(N)}_i,B^{(N)}_i$ modes are in fact band edge modes in their intervals of free propagation. In our case $L$ gives the characteristic length, which, as we will demonstrate below, also determines the mode frequency. As $N$ increases, the eigenfrequencies of the modes shift towards bandgap. The associated decrease of the local group velocity contributes to the increase of the Q-factor of the resonators and further reduction of the group velocity in $A^{(N)}_i,B^{(N)}_i$ bands in $N\rightarrow\infty$ limit.  

Eq. \ref{q_from_t} suggests that the dispersion relation $\omega(q)$ is independent of how segment of length $L$ (the period of our structure) is chosen. However, the transmission coefficient through $0<x<Na$ and $-Na/2<x<Na/2$ segments of the crystal show very different spectral composition, Fig. \ref{transmission}(a). In order to understand how these markedly different functions lead the same $\omega(q)$, we will analyze the phase of the transmission coefficient $\phi$, Fig. \ref{transmission}(b).

In a one-dimensional periodic system such as ours, the wave number $q(\omega)$ in Eq. \ref{q_from_t} is equal to the integrated density of electromagnetic states. It is by definition a monotonically increasing function of the frequency in extended Brillouin zone scheme. In PhSC, $q\times L$ increases by $\pi$ every time the frequency is increased through an allowed band, {\it c.f.} bold line in Fig. \ref{transmission}(b). At the frequency in the middle of the band $\cos(qL)=0$ because $q\times L=\pi\times (m+1/2)$. From Eq. (\ref{q_from_t}) one can see, that $\phi$ should be equal to $\pi\times (m+1/2)$ at the same frequency. In the finite system mode counting phase defined\cite{lifshitz} as $\tan(\widetilde{\phi})=E^{\prime}/E$ coincides with the phase of the transmission coefficient $\phi\equiv\widetilde{\phi}$ that explains the monotonic behavior of $\phi(\omega)$. Eq. \ref{q_from_t} leads to the fact that the quasi-states of the finite system occur at the same place as the corresponding band center of the lattice, irrespective of the definition of the unit cell. Thus, as it can be also seen from Fig. \ref{transmission}(b), $\phi(\omega)$ and $q(\omega)L$ intersect at $\pi\times (m+1/2)$. 

Taylor expansion of the phase around frequency at the center of a pass band, $\omega_0$, where $q(\omega)L=\pi\times (m+1/2)$ gives 
\begin{equation}
\cos(q(\omega)L)=(\omega-\omega_0)\times \frac{(-1)^m}{\left|t(\omega_0)\right|}\frac{d\phi(\omega_0)}{d\omega}.
\label{res_expansion}
\end{equation}  
Here, the term that contained $d\left|t(\omega_0)\right|/d\omega$ dropped out because $\cos(q(\omega_0)L)=0$. Comparing Eqs. (\ref{crow_band},\ref{res_expansion}) shows that $\left|t(\omega_0)\right|^{-1}d\phi(\omega_0)/d\omega$ determines $Q=1/\kappa$ and not only $\left|t(\omega_0)\right|$. Suppressed transmission compensates for slow phase change ({\it e.g.} solid line in Fig. \ref{transmission}(b) in the high frequency spectral region) and leads to the identical $q(\omega)$ for two different definitions of the unit cell.
 
We also note that if the segment is chosen such that the correspondent ``cavity'' is located in the geometrical center ($\left|t(\omega_0)\right|=1$), FWHM of the resonance ($\Gamma$) in the transmission coefficient is equal to the width of the pass band in the periodic lattice. This fact follows from Eqs. (\ref{q_from_t},\ref{lorentzian}). It further emphasizes the analogy with CROW structures that we explored in Ref. \cite{our_ol_crow}.

We conclude this section by noting that long-range refractive index modulation creates alternating spatial regions that can serve as resonators separated by the tunneling barriers. Hybridization of the cavity resonances creates a series of photonic bands with low dispersion. The envelope of the eigen-states in these bands $A^{(N)}_i,B^{(N)}_i$ is a slowly varying function of the coordinate, see Fig. \ref{intensity}. This effect stems from states proximity to photonic bandedge of the underlying single-periodic lattice. Possibility of separation of length scales into short ($a$ of rapid field oscillations) and long ($L$ of the slow amplitude variation) will inform our further analytical studies presented in the following sections. 

\section{Resonant Approximation}

Forbidden gaps in spectra of a periodic system arise due to resonant interaction of the wave with its Bragg-scattered counterpart\cite{ashcroft}. The scattered wave appears due to presence of Fourier harmonic in the spectrum of the periodic ``potential'', which is in the case of Helmholtz equation
\begin{equation}
E''(x) + \frac{\omega^{2}}{c^2}\delta\varepsilon(x) E(x) = \frac{\omega^{2}}{c^2}\overline{\varepsilon}E(x),
\label{helmholtz}
\end{equation}
is represented by $(\omega^{2}/c^2)\delta\varepsilon(x)\equiv (\omega^{2}/c^2)\left[\varepsilon(x)-\overline{\varepsilon}\right]$. Here we introduced the average value of the dielectric function $\overline{\varepsilon}=\overline{\varepsilon(x)}=\varepsilon_0+\Delta\varepsilon/[2(1+\gamma)]$. When $\gamma=0$, the condition $\Delta\varepsilon/\overline{\varepsilon}\ll 1$ is sufficient to obtain the position and width of spectral gaps. Otherwise, an additional condition $N\times\Delta\varepsilon/\overline{\varepsilon}\ll 1$ needs to be satisfied instead. We will discuss the physical meaning of this condition at the end of this section.

We begin by noticing that $\delta\varepsilon(x)$ of our choice (Eq. \ref{epsilon}) contains
\begin{equation}
\varepsilon(x)=\displaystyle\sum  _{m=-\infty}^{\infty}\varepsilon_{m}\exp\left[ i \frac{2\pi}{L}m x\right] 
\label{eps_fft}
\end{equation}
only eight nonzero Fourier harmonics: $m=\{\pm 1, \pm (N-1),\pm N, \pm (N+1))\}$. This fact allows one to do the exhaustive study of all resonant interactions as follows. Expressing $E(x)$ in terms of its Fourier components 
\begin{equation}
E(x)=\exp[iq(\omega )x]\displaystyle\sum  _{m=-\infty}^{\infty}E_{m}\exp\left[ i \frac{2\pi}{L}m x\right] 
\label{E_definition}
\end{equation}
leads to an infinite system of linear coupled equations
\begin{equation}
\left[ \frac{\omega^{2}}{c^{2}}\overline{\varepsilon}-\left(q(\omega )+\frac{2\pi}{L}m\right)^2\right]  E_{m} +\frac{\omega^{2}}{c^{2}}\displaystyle\sum_{m^{\prime}\neq 0} \varepsilon_{m^\prime}E_{m-m^{\prime}}=0,
\label{helmholtz_fourier}
\end{equation}
where $q$ is the Bloch number that varies in the first Brillouin zone $[0,\pi/L]$. For the extreme values of $q$ there exists spectral range where the term in brackets of Eq. \ref{helmholtz_fourier} can become simultaneously small for certain $m$ and $-m$ at $q=0$ and for $m$ and $-m-1$ at $q=\pi/L$. If $\varepsilon(x)$ contains harmonic $\varepsilon_{m^{\prime}}$ such that it couples these two Fourier components, the overall infinite system  Eq. \ref{helmholtz_fourier} can be reduced to two resonant equations.

\begin{widetext}
\begin{center}
\begin{table}
\begin{tabular}{lccc}
\hline
\hline
Resonant $q$,  (even $N$) & $\pi/L$  & $0$  & $\pi/L$ \\
\ \ \ \ \ \ \ \ \ \ \ \ \ \ \ \ \   (odd $N$) & $0$ & $\pi/L$ & $0$\\
Coupled components & $E_{-s},E_{N-(s+1)}$ & $E_{s-N},E_{s}$  &  $E_{-(s+1)},E_{N-s}$ \\
Coupling harmonics & $\varepsilon_{-(N-1)}$, $\varepsilon_{N-1}$ & $\varepsilon_{-N}$, $\varepsilon_{N}$ & $\varepsilon_{-(N+1)}$, $\varepsilon_{N+1}$ \\
Center frequency, $\omega_0$&$\displaystyle\frac{c\pi}{\sqrt{\overline{\varepsilon}}L}(N-1)$&$\displaystyle\frac{c\pi}{\sqrt{\overline{\varepsilon}}L}N$&$\displaystyle\frac{c\pi}{\sqrt{\overline{\varepsilon}}L}(N+1)$\\
Normalized width, $\Delta\omega/\omega_0$& $\varepsilon_{(N-1)}/\overline{\varepsilon}$ & $\varepsilon_{N}/\overline{\varepsilon}$ & $\varepsilon_{(N+1)}/\overline{\varepsilon}$ \\
\hline
\hline
\end{tabular}
\caption{\label{eps_expansion} Results of resonant approximation analysis of Eq. \ref{helmholtz_fourier} with dielectric function given by Eq. \ref{epsilon}. Three columns correspond to the three resonant photonic band gaps that appear in the spectrum of the dual-periodic PhSC. The expressions hold for both even and odd $N$ for the choice of parameter $s$: $N=2s$ and $N=2s+1$ respectively.}
\end{table}
\end{center}
\end{widetext}

\begin{figure}
\centerline{{\scalebox{0.4}{\includegraphics{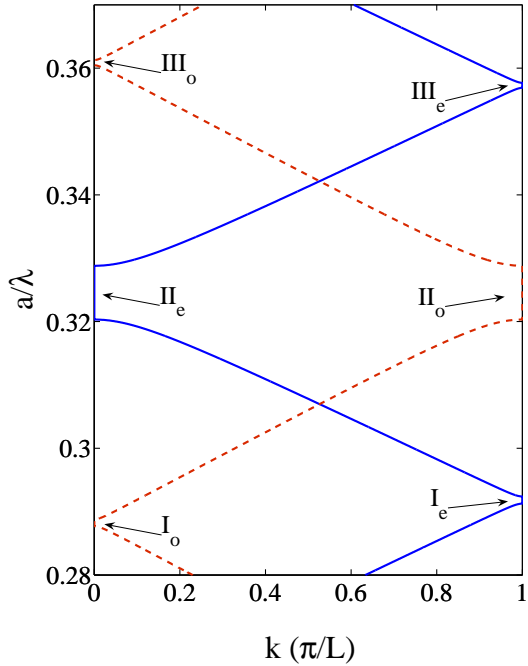}}}}
\caption{\label{perturbation} Dispersion relation computed with transfer matrix formalism for $\varepsilon_0=2.25$, $\Delta\varepsilon=0.32$, $N=9$ (dashed line) and  $N=10$ (solid line) and the modulation parameter $\gamma$ is equal to $0.25$. For this set of parameters, the applicability condition Eq. \ref{perturb_accuracy} of the resonant approximation is satisfied.}
\end{figure}

The results of such analysis are summarized in Table \ref{eps_expansion} and the correspondent band structure is shown in Fig. \ref{perturbation}. Introduction of the long range modulation in the dielectric constant results in expansion of the unit cell from $a$ to $L=Na$ and, thus, the reduction of the Brillouin zone accompanied by the folding of photonic bands. The cases of even $N=2s$ and odd $N=2s+1$ should be distinguished. In the former, the primary photonic bandgap (${\rm II_e}$) of the single-periodic lattice is reappears at $q=0$, whereas in the latter (${\rm II_o}$) it is located at  $q=\pi/L$. Our analysis shows that the nearest in frequency gaps, ${\rm I_{o,e}}$ and ${\rm III_{o,e}}$, also become resonant, Fig. \ref{perturbation}. For the considered refractive index modulation of Eq. \ref{epsilon}, the normalized width of the satellite gaps is smaller $\displaystyle\varepsilon_{(N-1)}/\overline{\varepsilon}=\frac{\gamma\Delta\varepsilon /4\overline{\varepsilon}}{1+\gamma}$ than that of the central gap by a factor of $\gamma$. This parameter by definition is less than unity. 

We can see that folding and the offset of the formation of flat $A^{(N)}_1$, $B^{(N)}_1$ bands is captured in this approximation. The criterion of its applicability  can be found by considering the contributions of non resonant terms in Eq. \ref{helmholtz_fourier}. We find that for all three gaps the criteria are qualitatively the same. Therefore, we present the detailed analysis of only one particular resonance, ${\rm III_e}$. The condition that the closest non-resonant Fourier components $E_{-s-2}$, $E_{-s}$, $E_{s-1}$ and $E_{s+1}$ are smaller than the resonant ones $E_{-s-1}$ and $E_{s}$ leads to
\begin{equation}
\displaystyle\frac{(N+1)^{2}(\varepsilon_{1}+\varepsilon_{N})}{4N\overline{\varepsilon}+2(N+1)^2\varepsilon_{N-1}}\ll 1.
\label{perturb_accuracy}
\end{equation}
In the limit of very large $N$ the second term in the denominator becomes dominant and this condition cannot be satisfied regardless of the value of $\Delta\varepsilon$. Thus, $N$ should be finite. It can be seen that the condition under which the first term in the denominator is dominant, is consistent with the entire inequality Eq. \ref{perturb_accuracy} and can be put to read $N\varepsilon_i/\overline{\varepsilon}\ll 1$. Taking the case most restrictive $\varepsilon_i$ we finally obtain
\begin{equation}
\displaystyle\frac{\Delta\varepsilon}{8\overline{\varepsilon}} \times N\ll 1,
\label{perturb_accuracy_simple}
\end{equation}
where we neglected by $\gamma$ for simplicity. 

The derived expression Eq. \ref{perturb_accuracy_simple} has a clear physical meaning. Indeed, from Table \ref{eps_expansion}, the frequency of the bandgaps ${\rm I}$ and ${\rm III}$ approach the central gap inherited from the single periodic system as $1/N$. At some point a bandgap of width $\Delta\omega/\omega_0=\varepsilon_{i}/\overline{\varepsilon}$ begins to substantially perturb the pass band of width $q_{max}\times c\simeq\omega_0/N$ separating the consecutive gaps. The resonant approximation breaks when these two scales become comparable. This condition results in Eq. \ref{perturb_accuracy_simple}. In other words, the approximation considered in this section can at most capture the onset of flattening trend in $A^{(N)}_1$, $B^{(N)}_1$ bands and fails when $N$ is increased to the degree when these states become abnormally flat -- $\Delta\omega/q_{max}\ll c/\overline{\varepsilon}$ throughout the band. More sophisticated approaches are considered below.

\section{Effective Medium Approximation}

Gratings written in the core of photosensitive optical fibers are often analyzed with the help of coupled-mode theory (CMT)\cite{coupled_mode_theory}. In both shallow gratings with long-range modulation in fibers\cite{1d_transfer_matrix1,bragg_grating} and our PhSC, the forward and backward propagating (locally) waves continuously scatter into each other. The advantage of CMT is in that it considers the amplitudes of the forward and backward waves directly. This tremendously simplifies Maxwell equations. Ref. \cite{desterke} also considered the fiber gratings with deep piece-wise constant index modulation. In this section we employ the CMT-based method developed by Sipe {\it et al} \cite{1d_transfer_matrix1} to obtain the spectral positions of the flat photonic bands formed in PhSC. 

For {\it shallow modulation}, small $\Delta\varepsilon$, our Eq. (\ref{epsilon}) can be brought to resemble the model function considered in Ref. \cite{1d_transfer_matrix1}
\begin{equation}
n(x)/n_0=1+\sigma(x)+2\kappa(x)\cos\left[2k_0 x+\varphi(x)\right]
\label{sipe_nofx}
\end{equation}
with the following choice of parameters
\begin{eqnarray}
\sigma(x)&=&\displaystyle\frac{\displaystyle\frac{\gamma\Delta\varepsilon/4}{1+\gamma}}{\varepsilon_0+\displaystyle\frac{\Delta\varepsilon /2}{1+\gamma}}\times\cos\frac{2 \pi}{L}x; \nonumber \\
\kappa(x)&=&\displaystyle\frac{\displaystyle\frac{\Delta\varepsilon/8}{1+\gamma}}{\varepsilon_0+\displaystyle\frac{\Delta\varepsilon /2}{1+\gamma}}\times\left(1+\gamma\cos\frac{2 \pi}{L}x\right);                     \label{sipe_parameters}\\
\varphi(x)&\equiv &0;\ \ \ \  n_0=\left(\varepsilon_0+\frac{\Delta\varepsilon /2}{1+\gamma}\right)^{1/2};\ \ \ \  k_0=\pi/a.\nonumber    
\end{eqnarray}
CMT of Ref. \cite{1d_transfer_matrix1} is applicable as long as these functions have {\it slow dependence on} $x$, on the scale much larger then $k_0^{-1}$. This condition is indeed satisfied in PhSC with $N\gg 1$.

\begin{figure}
\centerline{{\rotatebox{-90}{\scalebox{0.34}{\includegraphics{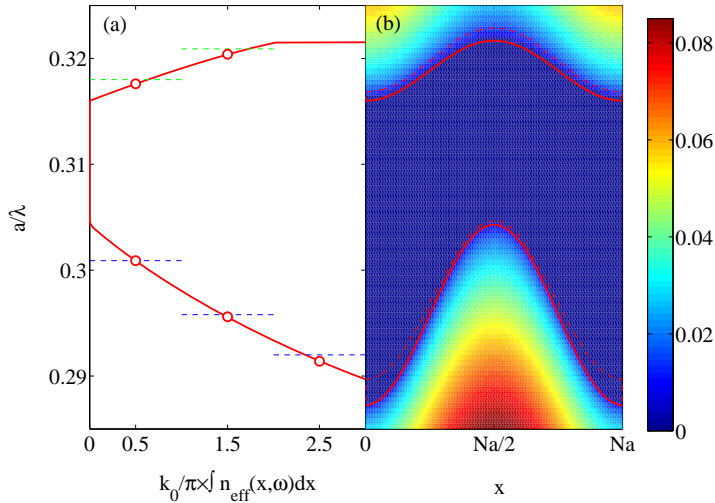}}}}}
\caption{\label{neff_quatization} (a) The value of integral in Eq. (\ref{wkb}), solid line, as a function of frequency is shown. For an easy comparison with (b), the plot is transposed so that $\omega$ is plotted along y-axis. The open circles depict frequencies that satisfy the quantization condition of Eq. (\ref{wkb}). The dashed lines denote the actual position of the photonic states, as determined by direct numerical analysis of Section 3. (b) Color-coded plot of $Re[n_{eff}(x,\omega)]$ given by Eq (\ref{neff}). The solid line shows the boundary of the region where $Im[n_{eff}(x,\omega)]\neq 0$. For comparison we also show the local PBG of Fig. \ref{nofx}(b), dashed line. In both (a) and (b), the parameters of Fig. \ref{nofx} are adopted.}
\end{figure}

By introducing small detuning parameter
\begin{equation}
\delta=\frac{\omega-\omega_{0}}{\omega_{0}}\ll 1,\ \ \ \omega_{0}=\frac{k_0c}{n_0}\nonumber
\end{equation}
we can, following Ref. \cite{1d_transfer_matrix1}, obtain the governing equation for the quantity $E_{eff}$ related to the envelope of the electric field
\begin{equation}  
\frac{d^{2}E_{eff}}{dx^{2}} +k_0^{2}n^{2}_{eff}(x,\omega)E_{eff}=0.   \label{e_eff}
\end{equation}
Frequency and position dependent effective refractive index 
\begin{equation}
n_{eff}=\lbrace ( \sigma(x)+\Delta)^{2}-\kappa(x)^{2}\rbrace^{1/2}
\label{neff}
\end{equation}
determines whether propagation is locally allowed (real $n_{eff}$) or forbidden (imaginary $n_{eff}$). This is similar to our definition of local PBG diagram which we studied numerically in Section 3 Fig. \ref{neff_quatization}b compares CMT's region of evanescent propagation (solid lines) to the numerical calculation (dashed lines). We attribute the relatively small discrepancy to the assumption of shallow modulation we made in arriving to Eq. (\ref{neff}).

Eq. (\ref{e_eff}) is formally similar to Schrodinger equation. Our previous analysis shows that the single-period states associated with photonic bands $A_i^{(N)},B_i^{(N)}$ are confined to the region of classically allowed propagation in the language of quantum mechanics. By analogy, WKB approach of quantum mechanics can be applied \cite{1d_transfer_matrix1} to determining quantization of energies inside our optical equivalent of a quantum well
\begin{equation}
k_{0}\displaystyle\int_{x_{L}}^{x_{R}}n_{eff}(x,\omega)dx=(m+1/2)\pi
\label{wkb}
\end{equation}
$x_{L}$ and $x_{R}$ are respectively the left and right turning points defined by the condition $n_{eff}(x_{L,R},\omega)=0$, $m$ is an integer. The solid line in Fig. \ref{neff_quatization}a depicts the value of the integral, x-axis, as function of $\omega$ obtained numerically. The open circles denote the the frequencies at which the quantization condition of Eq. (\ref{wkb}) is satisfied. In the system with the parameters we used for illustration in previous sections, the obtained solutions are in good agreement with numerical results obtained with transfer matrix approach, Section 3. This suggests that index variation given by $\Delta\varepsilon=1,\ \varepsilon_0=2.25$ was sufficiently small for this approach to be applicable.

We finish the current section by noting that it would be desirable to retain the attractive property of CMT envelope approach without being constrained by the condition of small refractive index modulation. The latter may not always be justified in the experimental situation of interest\cite{bertino_apl2004}. In the following section we develop such an approach.

\section{Bogolyubov-Mitropolsky Approach}


In this section we will consider the standing-wave solutions of Eq. (\ref{helmholtz}). In this case, the corresponding $E(x)$ can be chosen a real function by appropriate choice of normalization. Than, we make Bogolyubov anzatz \cite{landa,bogolyubov}:
\begin{eqnarray}
E(x)=&A(x)\cos\left(k_0 x+\phi(x)\right) \label{bogolyubov_anzatz}\\
dE(x)/dx=&-k_0A(x)\sin\left(k_0 x+\phi(x)\right), \nonumber
\end{eqnarray}
where as in the preceding section, $k_0=\pi/a$. The above equations defines the amplitude and phase functions. Its substitution into Eq. (\ref{helmholtz_fourier}) gives so-called Bogolyubov equations in the standard form\cite{landa,bogolyubov}
\begin{eqnarray}
\displaystyle\frac{d\phi(x)}{dx}&=&\frac{1}{k_0} \left[\frac{\omega^2}{c^2}\varepsilon(x)-k_0^2\right]\cos^2\left(k_0 x+\phi(x)\right) \label{bogolyubov_standard}\\
\displaystyle\frac{dA(x)}{dx}&=&\frac{A(x)}{2k_0} \left[\frac{\omega^2}{c^2}\varepsilon(x)-k_0^2\right]\sin2\left(k_0 x+\phi(x)\right), \nonumber
\end{eqnarray}
No approximations have been made so far. The structure of the above equation suggests that conditions $dA/dx\ll k_0A$ and $d\phi/dx\ll k_0\phi$ can be satisfied in the vicinity of the spectral region where $(1/k_0)\left[\omega^2/c^2\overline{\varepsilon(x)}-k_0^2\right]\ll k_0$. Overbar denotes the average over one period. Comparison with the analysis in the previous sections shows that this condition is satisfied in the vicinity of the primary photonic bandgap. In the system of interest $N\gg 1$, this observation justifies ``averaging-out'' the fast spectral components -- Mitropolsky technique \cite{bogolyubov}. The averaging procedure leads to the following system of nonlinear equations on the slow-varying amplitude and phase
\begin{widetext}
\begin{eqnarray}
\displaystyle\frac{d\phi(x)}{dx}&=&\frac{1}{2k_0} \left[\frac{\omega^2}{c^2}\varepsilon_{0}-k_0^2+\frac{\omega^2}{c^2}\frac{\Delta\varepsilon /2}{1+\gamma}\left(1+\gamma\cos\frac{2\pi}{L}x\right)\left(1+\frac{1}{2}\cos2\phi(x)\right)\right] \label{bogolyubov_phi}\\
\displaystyle\frac{d \log A(x)}{dx}&=&\frac{1}{2k_0}\frac{\omega^2}{c^2}\frac{\Delta\varepsilon /2}{1+\gamma}\left(1+\gamma\cos\frac{2\pi}{L}x\right)\sin2\phi(x). \label{bogolyubov_A}
\end{eqnarray}
\end{widetext}
In derivation of Eqs. (\ref{bogolyubov_phi},\ref{bogolyubov_A}) we used the explicit form of $\varepsilon(x)$ given by Eq. (\ref{epsilon}). 

\begin{figure}
\centerline{{\rotatebox{-0}{\scalebox{0.4}{\includegraphics{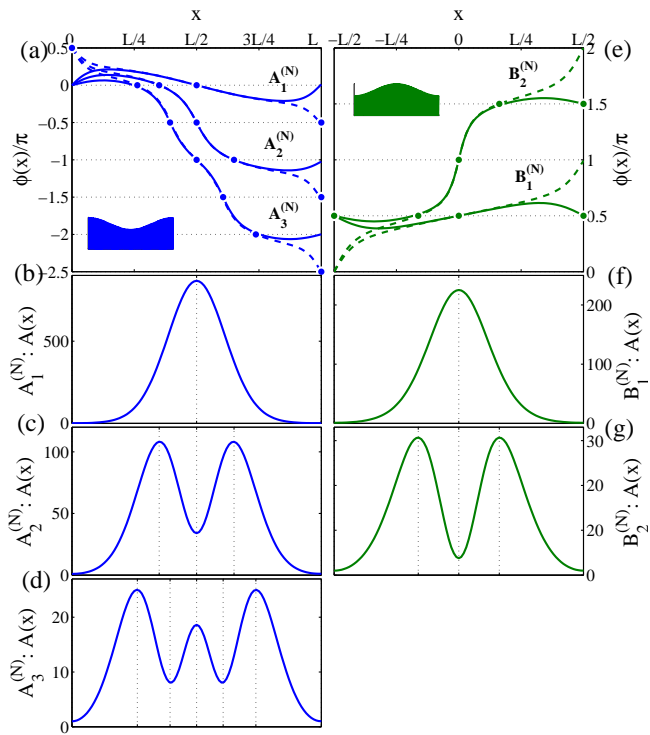}}}}}
\caption{\label{bmrk} Filled circles in panels a,e denote the spacial position where the particular $\phi(x)$ turns to $m\pi/2$. At these special points $dA(x)/dx=0$ denoted by the vertical dashed lines in b-d and f-g panels. }
\end{figure}

We begin the analysis of the obtained Eqs. (\ref{bogolyubov_phi},\ref{bogolyubov_A}) with a discussion of the appropriate boundary conditions. In deriving these equations we limited the consideration to the real-valued solutions of the original Eq. (\ref{helmholtz_fourier}), that can be found only for a discrete set of frequencies. At the sought frequencies, the corresponding amplitude function should incur the periodicity of the dielectric function Eq. (\ref{epsilon}) that implies
\begin{eqnarray}
\phi(L)=\phi(0)+m\pi \label{phiBC}\\
\sin 2\phi(0,L/2,L)=0 \label{ABC}.
\end{eqnarray}
The first condition is obtained by requiring $\sin2\phi(x)$ in Eq. (\ref{bogolyubov_A}) to be periodic. Symmetry of the modulation profile $A(x)$, see Fig. \ref{nofx}(a), and continuity of its derivative lead to $dA(x=0,L/2,L)/dx=0$ condition. This can only be satisfied by requiring Eq. (\ref{ABC}), because other factors on the right hand side of Eq. (\ref{bogolyubov_A}) are positive functions.

The equation which determines evolution of the phase, Eq. (\ref{bogolyubov_phi}), is self-contained. Hence, its solution together with the constraints given by Eqs. (\ref{phiBC},\ref{ABC}) is sufficient to obtain the spectrum of system and $\phi(x)$. The amplitude is to be recovered in the second step by simple integration of Eq. (\ref{bogolyubov_A}) with the found phase $\phi(x)$.


Fig. \ref{bmrk} shows the solutions of the Eqs. (\ref{bogolyubov_phi},\ref{bogolyubov_A},\ref{phiBC},\ref{ABC}) obtained by fourth order Runge-Kutta numerical method. In accord with our expectation, for each band there exist two solutions $\phi(x)$ which correspond to the standing-wave bandedge modes at $k=0,\pi/L$, Fig. \ref{bmrk}a,e. The corresponding solutions of the amplitude equation, Fig. \ref{bmrk}b-d,f-g, agree with the envelopes extracted from the direct solutions of the Helmholtz equation, Fig. \ref{intensity}. The eigenvalues of Eq. (\ref{bogolyubov_phi}) also give the frequencies that correspond to the bandedge states that are also in excellent agreement (the observed deviation is less then $0.1\%$), Fig. \ref{intensity}. Knowlendge of the bandedge frequencies allows to determine all parameters of the tight-binding approximation for $\omega(k)$, Eq. (\ref{dispersion}). Therefore, the entire band structure in the spectral region of the flat bands can be obtained solely from the solutions of the amplitude-phase equation.

Filled circles in panels a,e of Fig. \ref{bmrk} denote the spatial position where the particular $\phi(x)$ turns to $m\pi/2$. At these special points $dA(x)/dx=0$ denoted by the vertical dashed lines in b-d and f-g panels. Thus, the overall phase accumulated by $\phi(x)$ over one period, is an important parameter indicative of the spatial structure of the amplitude. At the bandedge frequencies of the bands $A_1^{(N)}$ and $B_1^{(N)}$ (see Section 3 for notations), the phase is a bound function $|\phi(x)-\phi(0)|\le\pi/2$. Therefore, $x=0,L/2,L$ are the only positions where the corresponding amplitude function takes minimum/maximum values, Fig. \ref{bmrk}b,f, -- $A(X)$ has only one ``hump'' for $A_1^{(N)}$ and $B_1^{(N)}$. Comparison of $\phi(x)$ computed for $k=0$ (solid lines) and $k=\pi/L$ (dashed lines) edges of each photonic band shows (Fig. \ref{bmrk}) that the difference occurs in the spacial regions where the electromagnetic waves propagate via tunneling mechanism in language of CROWs of Section 3. In these regions $A(x)$ is small, which explains the small spectral width of the corresponding photonic bands.

As the eigen-frequencies of the higher order states $A_2^{(N)},B_2^{(N)}...$ shift further away from the primary band-gap region, Fig. \ref{nofx}b, $\phi(x)$ becomes progressively steeper function, leading to the steady increase of the number of ``humps'' in $A(x)$, \ref{bmrk}. Such progression accelerates the spacial dependence of amplitude and eventual breakdown of the scale separation approximation used in the derivation of Eqs. (\ref{bogolyubov_phi},\ref{bogolyubov_A}). Nevertheless, such loss of applicability occurs well outside of spectral region of interest, demonstrating the robustness of the developed approach.

\section{Summary}

In this work we studied the optical properties of dual-periodic photonic superlattice with four different theoretical methods. Although each method has its limitations, the results obtained in framework of each model complement each other.

Numerical simulation with transfer matrices yields the direct result for the photonic band structure. This method, however, provides little physical insight into the nature of the photonic bands. In Section 3 we compared the spectrum of the infinite (periodic) crystal with the transmission spectrum of the finite system with the length equal to one period of the superstructure. We identified the individual transmission resonances with the photonic bands and found one-to-one correspondence. Furthermore, the spatial distribution of the fields at a resonance demonstrated that in $L\gg a$ limit the envelope (amplitude) of the state changes slowly -- on the scale of $L$. 

With a method commonly employed in condensed matter physics we investigated the resonant interactions between Bloch waves when the second, longer-scale, modulation is introduced. We showed that flattening of the photonic bands is related to, but goes beyond band folding. It arises due to the increasing coupling between Bloch waves with k-vectors at the boundaries of Brillouin zone. The subsequent increase of the band-gaps regions ``squeezes'' the the bands making them progressively flatter as $N=L/a$ is increased. Although, as showed in the Section 4, this approach fails for very large $N$, it still provided an insight into the origin of the anomalously small dispersion in spectra of PhSC.
 
Diffraction gratings introduced in optical fibers are often spatially modulated. Coupled-mode theory has been developed to reduce the problem to a study of the amplitudes of the forward and backward propagating waves and to avoid the direct solution of Maxwell equations. Although the refractive index contrast induced in optical fibers is orders of magnitude smaller then in the systems we are concerned with in this work, CMT-based approach of Section 5 provided a clear physical picture. It showed that the electromagnetic states of our optical resonators can be thought of as the eigen-states of the photonic wells. This further reinforced the analogy with CROWs that we developed in Section 3.

We proceeded by noting a formal similarity between the Helmholtz equation with the considered dielectric function and the equation describing parametric resonance in oscillation theory. Adopting amplitude-phase formalism accompanied by separation of scales (short $a$ and long $L=N\times a$) allowed us to derive to tractable set of equations on the envelope functions. This enabled the study physically meaningful mode profiles directly, without assuming smallness of modulations of refractive index. 

We find that dual-periodic photonic lattice is a promising system for studying slow-light phenomena. It can be viewed as an array of evanescently-coupled optical cavities. The cavities are formed due to presence of the second, long-range, modulation which creates alternating spatial regions of allowed and forbidden propagation. Thus, analogies with multiple-quantum-well and coupled resonant optical arrays (CROWs) are appropriate. Unlike other CROW implementations, the studies structures can be produced with holography-based approach, which enables fabrication of arrays with a large number of identical resonators. The latter is crucial condition from an experimental point of view. It should ensure that the resonances of the individual cavities couple to form flat photonic bands. 

\section{Acknowledgments}

AY acknowledges support from University of Missouri-Rolla. MH acknowledges the support of University of Missouri-Rolla Opportunities for Undergraduate Research Experiences (UMR-OURE) scholarship and Milton Chang Travel Award by the Optical Society of America.


\end{document}